\documentclass[conference]{IEEEtran}
\IEEEoverridecommandlockouts
\usepackage{cite}
\usepackage{amsmath,amssymb,amsfonts}
\usepackage{algorithmic}
\usepackage{graphicx}
\usepackage{textcomp}
\usepackage{xcolor}
\usepackage{booktabs} 
\usepackage{multirow}
\usepackage{makecell}
\usepackage{hyperref} 
\usepackage{fancyhdr} 

\def\BibTeX{{\rm B\kern-.05em{\sc i\kern-.025em b}\kern-.08em
    T\kern-.1667em\lower.7ex\hbox{E}\kern-.125emX}}
\begin{document}

\title{Leveraging Multimodal Methods and Spontaneous Speech for Alzheimer's Disease Identification \\
% {\footnotesize \textsuperscript{*}Note: Sub-titles are not captured for https://ieeexplore.ieee.org  and
% should not be used}
}

\author{
\IEEEauthorblockN{ Yifan Gao\textsuperscript{†}, Long Guo\textsuperscript{†},  Hong Liu*}
\IEEEauthorblockA{
\textit{College of Computer Science } \\
\textit{National Key Laboratory of Fundamental Science on Synthetic Vision} \\
\textit{Sichuan University}\\
Chengdu, China \\
}
\thanks{\textsuperscript{†}These authors contributed equally to this work.}
\thanks{\textsuperscript{*}Corresponding author}
}
\maketitle

\begin{abstract}

Cognitive impairment detection through spontaneous speech is a promising avenue for early diagnosis of Alzheimer’s disease (AD) and mild cognitive impairment (MCI), where timely intervention can significantly improve patient outcomes. The PROCESS Grand Challenge at ICASSP 2025 addresses these tasks by promoting innovative classification and regression methods for detecting cognitive decline. In this paper, we propose a multimodal fusion strategy that combines interpretable linguistic features with temporal embeddings extracted from pre-trained models. Our approach achieves an F1-score of 0.649 for the classification task (predicting healthy, MCI, dementia) and an RMSE of 2.628 for the regression task (MMSE score prediction), securing the top overall ranking in the competition.

\end{abstract}


\section{Introduction}
Alzheimer’s disease (AD) is a progressive neurodegenerative disorder that impairs language, speech, and memory \cite{AD_treatment}. Although incurable, timely intervention can slow its progression. Existing speech-based methods typically extract acoustic and linguistic features from spontaneous speech for automated diagnosis \cite{feature1}.

The PROCESS Grand Challenge (ICASSP 2025) aims to detect mild cognitive impairment (MCI) and dementia via classification (healthy, MCI, dementia) and regression (MMSE). Beyond the classic “Cookie Theft” description, PROCESS adds semantic and phonetic fluency tasks to enhance early detection.

\par
This paper presents our submission to the PROCESS Grand Challenge, combining interpretable semantic and temporal audio features via multimodal fusion. Our method achieved an F1-score of 0.649 and RMSE of 2.628, securing the top overall ranking in the competition.

\section{Methodology}

\subsection{Datasets}

The PROCESS dataset supports early-stage dementia detection, with each sample consisting of three types of audio tasks: Semantic Fluency Task (SFT), Phonemic Fluency Task (PFT), and Cookie Theft Description (CTD). SFT involves naming as many animals as possible within one minute, while PFT requires participants to generate as many words as possible starting with the letter 'P' within one minute. CTD captures a spoken description of the Cookie Theft picture\cite{cookie_theft}. Each sample includes diagnostic labels, gender, age, and, for some, MMSE scores. The dataset consists of 82 healthy controls (HC), 59 mild cognitive impairment (MCI) cases, and 16 dementia cases, offering a robust foundation for classification and prediction tasks.
\subsection{Preprocessing}
We imputed missing MMSE scores using category-wise means and refined them with a generative model. Next, Whisper-large automatically transcribed the audio into text (ASR). Finally, Koala's denoising model enhanced speech and suppressed noise, creating a “suppressed” version,\footnote{Koala: \href{https://picovoice.ai/docs/api/koala-web/}{https://picovoice.ai/docs/api/koala-web/}} 
while the Falcon model identified the primary speaker and concatenated their segments, generating a “subtracted” version.\footnote{Falcon: \href{https://picovoice.ai/platform/falcon/}{https://picovoice.ai/platform/falcon/}}

\subsection{Features extraction and fusion}
\subsubsection{Acoustic Features}
Acoustic features are made up of two parts: whisper embeddings and times whisper embeddings.

\textbf{Whisper Embedding}: 

The Whisper-large model uses a Transformer-based encoder to encode input audio signals into high-dimensional contextual feature vectors, capturing diverse characteristics\cite{2022_whisper}. In this study, audio was divided into 30-second segments, and features were extracted from each segment using the encoder. The segment-level features were then averaged to represent the entire audio, which was subsequently used for classification.

\textbf{Times Whisper Embedding}:

Dividing audio into 30-second segments captures only global features, thereby neglecting finer temporal information. To address this, we split the audio into 16 smaller chunks based on the 16,000 Hz sampling rate. Whisper embeddings from each chunk were concatenated to create a time-series-aware representation, referred to as Times Whisper Embedding, minimizing the impact of padding.

\subsubsection{Linguistic Features}
Language features are made up of three parts: general features, verbal fluency features, and pause features. 

\textbf{General Features}: The data were extracted using the Linguistic Feature Toolkit\footnote{LFTK is available at: \href{https://github.com/brucewlee/lftk}{https://github.com/brucewlee/lftk}}, a commonly used Python package in computational linguistics, which calculates 129 features. These features are used to measure the lexical richness and syntactic complexity of the text.

\textbf{Verbal Fluency Features}: For the PFT task, the number of correctly described words starting with the letter "P" is counted. For the SFT task, the number of correctly described animals is counted.

\textbf{Pause Features}: Pause features include duration-related descriptors extracted using a Voice Activity Detection (VAD)\cite{best2024} algorithm based on energy. It includes 16 descriptors used to describe the lengths of pauses and speech intervals.

\subsubsection{Multimodal fusion}
For classification tasks, we use a majority voting approach. Initially, each individual model is evaluated on its respective task using the validation set. Subsequently, we select the top-performing models and perform majority voting to determine the optimal combination based on validation set performance. In contrast, for regression tasks, we utilize an averaging voting method.

\section{Experiments and Results}

\begin{table}[htbp]
\begin{center}
\renewcommand{\arraystretch}{1.1} 
\textbf{TABLE I} \quad \textbf{Experiment Results of Cognitive Decline Detection}
\vspace{0.3cm} 

\begin{tabular}{|c|c|c|c|c|}
\hline
\textbf{Feature}  & \textbf{Task}   & \textbf{Model}   & \textbf{Val}  & \textbf{Test} \\
\hline
Baseline\cite{baseline} &  CTD   & \textbf{SVC}        & 0.390  &0.550   \\
\hline
\multirow{2}{*}{ Whisper Embedding }    
 &  PFT   & $A_1$.\textbf{Logistic}  & 0.492  &-   \\
\cline{2-5}
 &  SFT   & $A_2$.\textbf{RF}        & 0.467  & -   \\
\hline
\multirow{2}{*}{\makecell{Times \\ Whisper Embedding}} 
 &  CTD   & $A_3$.\textbf{MLP}    & 0.455  & -   \\
\cline{2-5}
 &  SFT   & $A_4$.\textbf{MLP}    & 0.513  & -   \\
\hline
\multirow{2}{*}{\makecell{Interpretable \\ Linguistic Feature}} 
 &  CTD   & $A_5$.\textbf{RF}        & 0.464  &-   \\
\cline{2-5}
 &  PFT   & $A_6$.\textbf{RF}        & 0.512  &-   \\
\hline
\multicolumn{3}{|c|}{MV$_1$$(A_1,A_2,A_5,A_6)$}   &  0.607  & 0.603  \\
\hline
\multicolumn{3}{|c|}{MV$_2$$(A_1,A_2,A_4,A_5,A_6)$}  &  0.618  &0.633  \\
\hline
\multicolumn{3}{|c|}{MV$_3$$(A_1,A_2,A_3,A_4,A_5,A_6)$}  &  0.627  &\textbf{0.649} \\
\hline
\end{tabular}

% 在这里增加说明
\vspace{0.5em}
\footnotesize
\raggedright
\textit{Note: $A_1, A_2, \dots$ are the model labels used in the voting approach; MV represents Majority Voting.}

\label{tab1}
\end{center}
\end{table}

\begin{table}[htbp]
\begin{center}
\renewcommand{\arraystretch}{1.1} 
\textbf{TABLE II} \quad \textbf{Experiment Results of MMSE Regression}
\vspace{0.3cm} 

\begin{tabular}{|c|c|c|c|c|}
\hline
\textbf{Feature}  & \textbf{Task}   & \textbf{Model}   & \textbf{Val}  & \textbf{Test} \\
\hline
Baseline\cite{baseline} &  CTD,PFT,SFT   & \textbf{RoBERTa}     & 3.110  & 2.985   \\
\hline
\multirow{2}{*}{ Whisper Embedding } 
    &  PFT   & $M_1$.\textbf{Logistic}     & 2.951  & -   \\
\cline{2-5}
    &  SFT   & $M_2$.\textbf{RF}         & 3.196  & -   \\
\hline
\multirow{2}{*}{\makecell{Times \\ Whisper Embedding}} 
    &  PFT   & $M_3$.\textbf{MLP}         & 3.257  & -   \\
\cline{2-5}
    &  CTD   & $M_4$.\textbf{MLP}         & 3.165  & -   \\
\hline
\multirow{2}{*}{\makecell{Interpretable \\ Linguistic Feature}} 
    &  PFT   & $M_5$.\textbf{RF}         & 2.973  & -   \\
\cline{2-5}
    &  CTD   & $M_6$.\textbf{RF}         & 2.886  & 4.655   \\
\hline
\multicolumn{3}{|c|}{AV$_1$($M_1,M_2,M_3$)}    & 2.585  & \textbf{2.628} \\
\hline
\multicolumn{3}{|c|}{AV$_2$($M_1,M_2,M_3,M_5$)}   & 2.706  & 2.723  \\
\hline
\end{tabular}

\vspace{0.5em}
\footnotesize
\flushleft
\textit{Note: $M_1, M_2, \dots$ are model labels  used in the voting approach; AV represents Averaging Voting.}

\label{tab2}
\end{center}
\end{table}

\subsection{Experiment setup}

The machine learning models use the default configurations of the classifiers. For the MLP method, 10 random seeds are generated. Each model is trained with a batch size of 32 for a total of 65 epochs, and the best model is recorded and saved.

The validation of our experiments followed a bootstrapping strategy (75$\%$ training, 25$\%$ validation), where the process was repeated 100 times to ensure variability in the different sets. 

\subsection{Experiment results}
Tables 1 and 2 show the results of cognitive decline detection and MMSE score prediction tasks respectively. The Whisper model and linguistic features performed well in different tasks. This demonstrates that both acoustic models and linguistic models can accurately distinguish between Alzheimer's patients and normal individuals.
\par
After applying voting method, the model's performance improved significantly. The best model achieved 0.649 F1 score in cognitive decline detection and an RMSE of 2.628 in MMSE prediction. 
\par

\section{Conclusion}
\par
In this paper, we present our submission for the PROCESS challenge. Our approach outperforms all other models in overall performance and maintains consistent performance across different dataset splits. The innovation in our work lies in the seamless integration of acoustic features and interpretable linguistic features, enhancing the efficacy of Alzheimer's disease detection.

\section{Acknowledgement}
\par
This work is partly supported by the National Natural Science Foundation of China under grant U20A20161 and the Fundamental Research Funds for Central Universities under grants SCU2024D055 and SCU2024D059.

\bibliographystyle{IEEEtran}
\bibliography{process.bib}

\end{document}